\documentclass{webofc}
\usepackage[varg]{txfonts}   
\usepackage{graphicx} 
\usepackage{url}

\begin{document}
\title{Adoption of a token-based authentication model for the CMS Submission Infrastructure}

\author{
\firstname{Antonio} \lastname{Pérez-Calero Yzquierdo}\inst{1,2}\thanks{\email{aperez@pic.es}} \and
\firstname{Marco} \lastname{Mascheroni}\inst{3}\thanks{\email{marco.mascheroni@cern.ch}} \and
\firstname{Edita} \lastname{Kizinevic}\inst{4} \and
\firstname{Farrukh Aftab} \lastname{Khan}\inst{5} \and
\firstname{Hyunwoo} \lastname{Kim}\inst{5} \and
\firstname{Maria} \lastname{Acosta Flechas}\inst{5} \and
\firstname{Nikos} \lastname{Tsipinakis}\inst{4} \and
\firstname{Saqib} \lastname{Haleem}\inst{6} \and
\firstname{Frank} \lastname{Würthwein}\inst{3}
on behalf of the CMS Collaboration}

\institute{
Centro de Investigaciones Energ\'eticas, Medioambientales y Tecnol\'ogicas (CIEMAT), Madrid, Spain \and
Port d'Informaci\'o Cientifica (PIC), Barcelona, Spain \and
University of California San Diego, La Jolla, CA, USA \and
European Organization for Nuclear Research (CERN), Geneva, Switzerland \and
Fermi National Accelerator Laboratory, Batavia, IL, USA \and
National Centre for Physics, Islamabad, Pakistan
         }

\abstract{%
The CMS Submission Infrastructure (SI) is the main computing resource provisioning system for CMS workloads. A number of HTCondor pools are employed to manage this infrastructure, which aggregates geographically distributed resources from the WLCG and other providers. Historically, the model of authentication among the diverse components of this infrastructure has relied on the Grid Security Infrastructure (GSI), based on identities and X509 certificates. In contrast, commonly used modern authentication standards are based on capabilities and tokens. The WLCG has identified this trend and aims at a transparent replacement of GSI for all its workload management, data transfer and storage access operations, to be completed during the current LHC Run 3. As part of this effort, and within the context of CMS computing, the Submission Infrastructure group is in the process of phasing out the GSI part of its authentication layers, in favor of IDTokens and Scitokens. The use of tokens is already well integrated into the HTCondor Software Suite, which has allowed us to fully migrate the authentication between internal components of SI. Additionally, recent versions of the HTCondor-CE support tokens as well, enabling CMS resource requests to Grid sites employing this CE technology to be granted by means of token exchange. After a rollout campaign to sites, successfully completed by the third quarter of 2022, the totality of HTCondor CEs in use by CMS are already receiving Scitoken-based pilot jobs. On the ARC CE side, a parallel campaign was launched to foster the adoption of the REST interface at CMS sites (required to enable token-based job submission via HTCondor-G), which is nearing completion as well. In this contribution, the newly adopted authentication model will be described. We will then report on the migration status and final steps towards complete GSI phase out in the CMS SI.
}
\maketitle
\section{Introduction}
\label{intro}

The CMS Submission Infrastructure (SI) is a critical component of the Compact Muon Solenoid (CMS) experiment at CERN~\cite{cms}, serving as the main computing resource provisioning system for CMS workloads. To effectively manage this infrastructure, a number of HTCondor pools are utilized~\cite{htcondor}, which aggregate resources from the Worldwide LHC Computing Grid (WLCG) and other providers~\cite{wlcg}.

Traditionally, the authentication model employed within the CMS SI has relied on the Grid Security Infrastructure (GSI), which is based on identities and X509 certificates. However, in line with the evolving standards and best practices in the computing field, there has been a growing shift towards token-based authentication models, and standards like OAuth, the industry-standard protocol for authorization.

Tokens offer several advantages over traditional identity-based authentication. Finer control over communication between entities is possible thanks to the lightweight nature and self-containedness of tokens. In fact, multiple tokens with different dedicated capabilities are used for different systems, in contrast with using one proxy for multiple purposes. Moreover, tokens allow for better adherence to the principle of least privilege, minimizing the impact in case of credential leaks.

Recognizing this trend, the WLCG has embarked on a comprehensive campaign to replace GSI with token-based authentication for all its workload management, data transfer, and storage access operations~\cite{wlcgtimeline}. As part of this broader effort, the CMS Submission Infrastructure group is actively phasing out the GSI component of its authentication layers and transitioning to IDTokens and Scitokens.

The adoption of token-based authentication within the CMS SI has been facilitated by the seamless integration of tokens into the HTCondor Software Suite, which forms the backbone of the infrastructure. This integration has allowed for a smooth migration of authentication between internal components of the SI. Furthermore, recent versions of the HTCondor-CE\cite{htcondorce} now support tokens as well, enabling CMS resource requests to be granted through token exchange at Grid sites utilizing this CE technology.

To ensure a successful transition, a comprehensive rollout campaign was conducted, which was completed by the third quarter of 2022. As a result, all HTCondor CEs utilized by CMS are now capable of receiving Scitoken-based pilot jobs. Additionally, a parallel campaign was launched to promote the adoption of the ARC-CE REST interface at CMS sites, which is a prerequisite for enabling token-based job submission. This campaign is nearing completion, further solidifying the integration of token authentication within the CMS SI.

In this paper, we will provide a detailed description of the newly adopted token-based authentication model within the CMS SI. We will also report on the migration status and outline the final steps towards the complete phase-out of GSI authentication within the CMS SI. By embracing token-based authentication, the CMS SI aims to enhance security, improve communication between entities, and align with the evolving standards and practices in the field of authentication and authorization.

\section{Using tokens for the internal components of the CMS Submission Infrastructure}
\label{sec-Internal}
\begin{figure}[ht]
\begin{center}
\includegraphics[width=8cm]{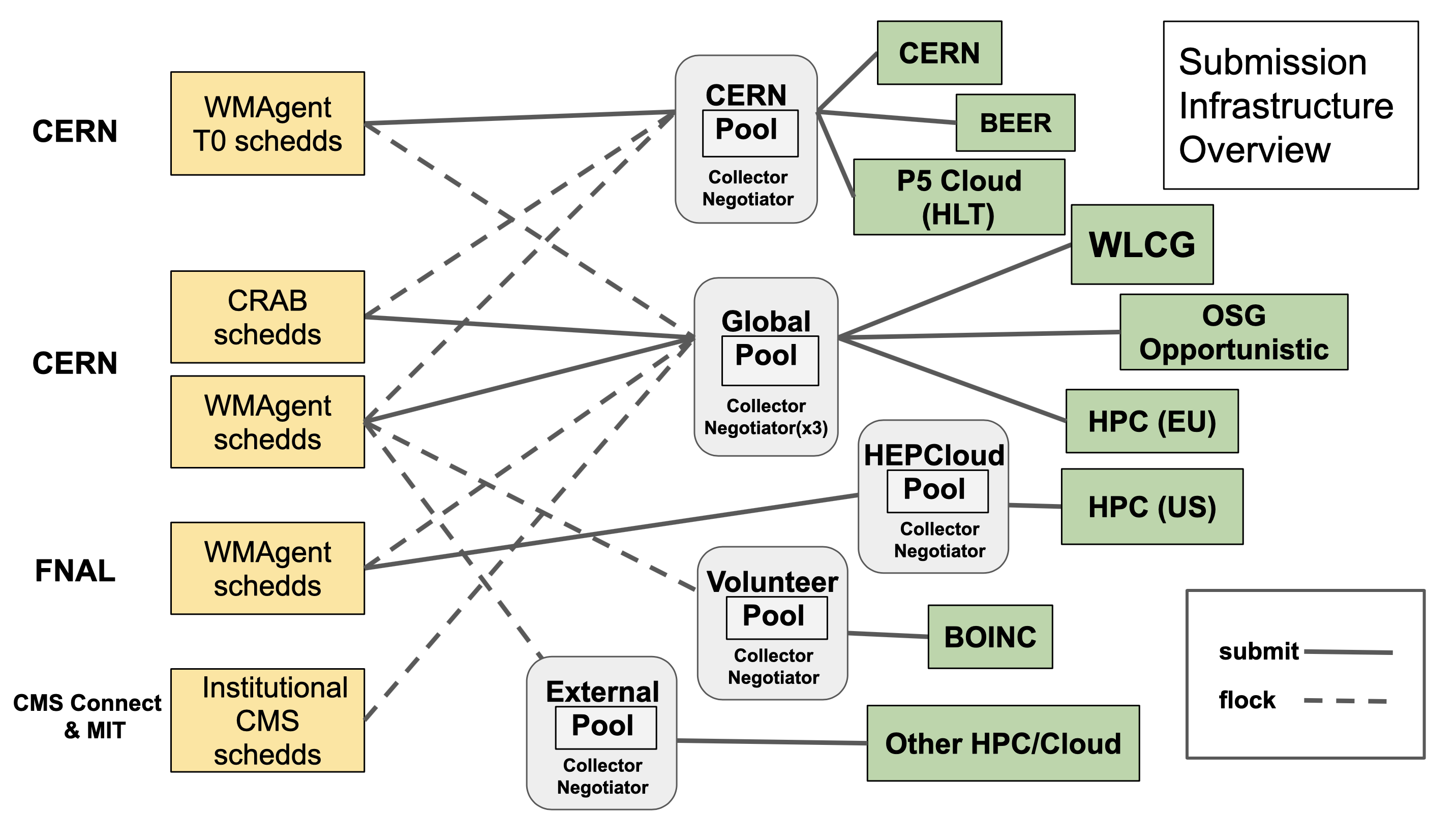}
\caption{CMS SI current configuration, including multiple federated pools of resources allocated from diverse origins (green boxes) and sets of distributed job schedulers (schedds), handling the Tier 0 and centralized production workloads (WMAgent), as well as analysis job submission (CRAB).} 
\label{fig:complexity}
\end{center}
\end{figure}

The CMS Submission Infrastructure gather resources from different providers using a set of federated HTCondor pools (see Figure \ref{fig:complexity})~\cite{sicomplexity}. For example, the CERN pool aggregates resources located at CERN, and the HEPCloud pool is used for HPC resources. Notably, the CMS Global pool stands out as it encompasses resources from the Grid, offering the largest resource pool for CMS workloads~\cite{globalpool}.

\begin{figure}[hb]
\begin{center}
\includegraphics[width=8cm]{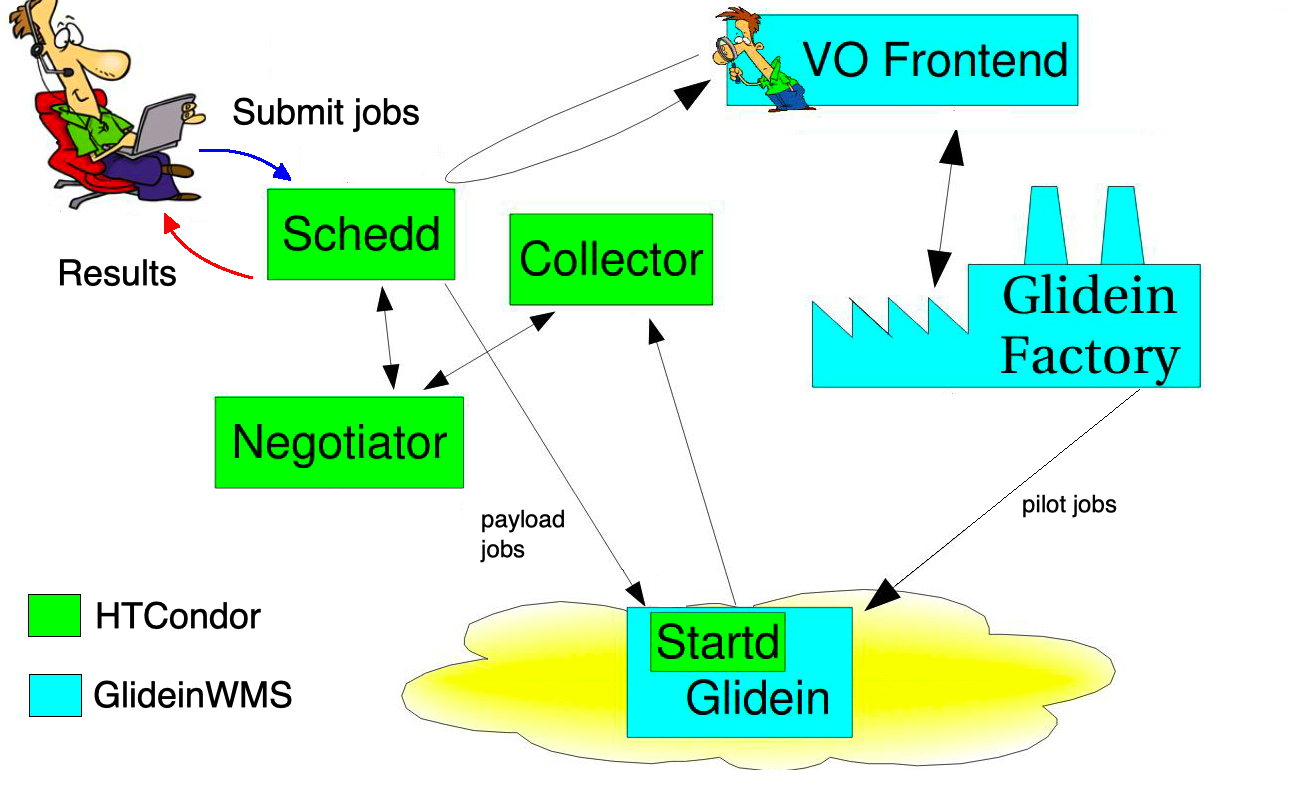}
\end{center}
\caption{GlideinWMS and HTCondor components building a dynamically sized pool of compute resources.}
\label{fig:pilots}
\end{figure}

The majority of these resource pools are built by using the glideinWMS workload resource management system~\cite{gwms}. As illustrated in Figure \ref{fig:pilots}, the glideinWMS tool uses a \emph{pilot job} approach~\cite{pilots}. A pivotal component, known as glideinWMS Frontend, monitors the job requests within the HTCondor schedulers. Subsequently, it dispatches pilot job requests to the factory, which, in turn, submits these requests to various Compute Elements (CEs) distributed worldwide. This process ensures that a pilot job can initiate on a local worker node at the respective site. Once initiated, the pilot launches a startd process, which connects to the resource pool, allowing the scheduling of user jobs.

The various components within this system operate on distinct hosts and require robust authentication and authorization mechanisms to facilitate secure communication and action execution. A lot of effort has been spent by the CMS Submission Infrastructure group in the migration from the GSI authentication and authorization to IDTokens.

\begin{figure}
\begin{center}
\includegraphics[width=12cm]{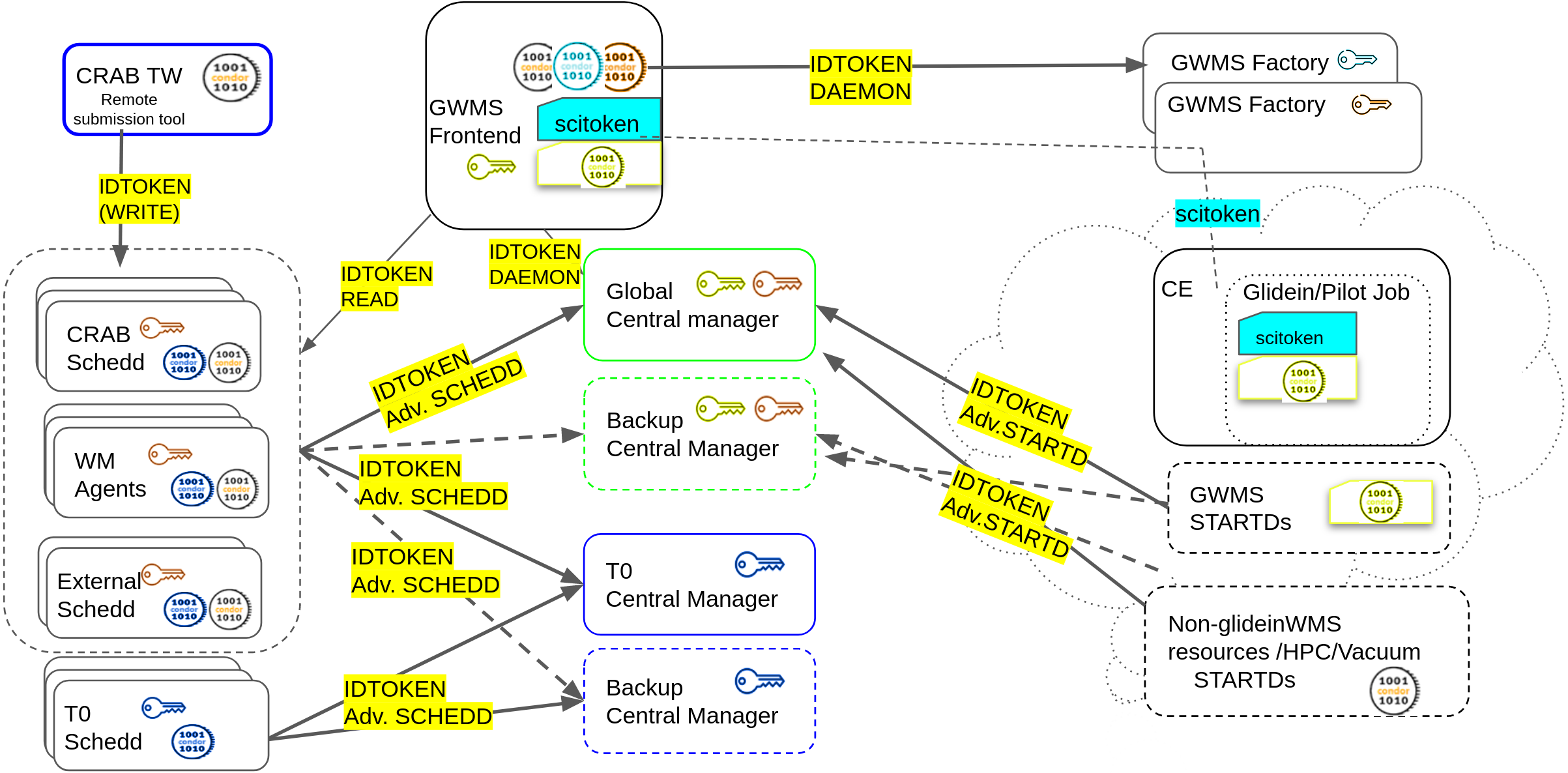}
\end{center}
\caption{The CMS Global Pool components, and where tokens are used for authentication as described in section \ref{sec-Internal}.}
\label{fig:SI_internals}
\end{figure}

IDTokens, a recent addition to HTCondor, introduce a native authentication mechanism designed to enhance security and identity verification within the HTCondor system. These tokens employ JSON Web Tokens (JWTs) and are signed with a symmetric key, ensuring data integrity and authenticity. Tokens can be generated conveniently using the HTCondor command-line interface, even remotely. Figure \ref{fig:SI_internals} describes how the new authorization system works.

Prior to 2022, the authorization process between the Workload Management (WM) systems, like WMAgent and Tier0, and the HTCondor schedulers relied on local file system (FS) authentication. Tokens were provided to the WM clients running on the schedulers to access the HTCondor daemon. In the case of CRAB, remote authentication was used and user proxies played a pivotal role. However, a notable shift has occurred towards a more streamlined approach with the adoption of a single service IDToken. Accommodating this change required development efforts within CRAB to transition from the previous practice of mapping multiple users with Argus. Since Spring 2022, authentication of Condor daemons has also undergone a significant evolution, with the entire Global pool now adopting IDTokens as the primary method for verifying the authenticity of daemons. A significant update occurred with the installation of HTCondor version 10 in early April, eliminating the GSI fallback. Tokens used by startd daemons on the worker node are transferred by GlideinWMS, which also undergone significant development in order to deal with tokens. 

The employment of multiple keys and tokens further bolster security. This strategy serves as a measure to mitigate potential impacts in the event of security incidents. By diversifying keys and tokens, the system can better isolate and contain security breaches, enhancing the overall resilience and security posture of the HTCondor daemons in the Global pool.

\section{Using tokens to access Grid resources}
\label{sec-External}

\begin{figure}
\begin{center}
\includegraphics[width=8cm]{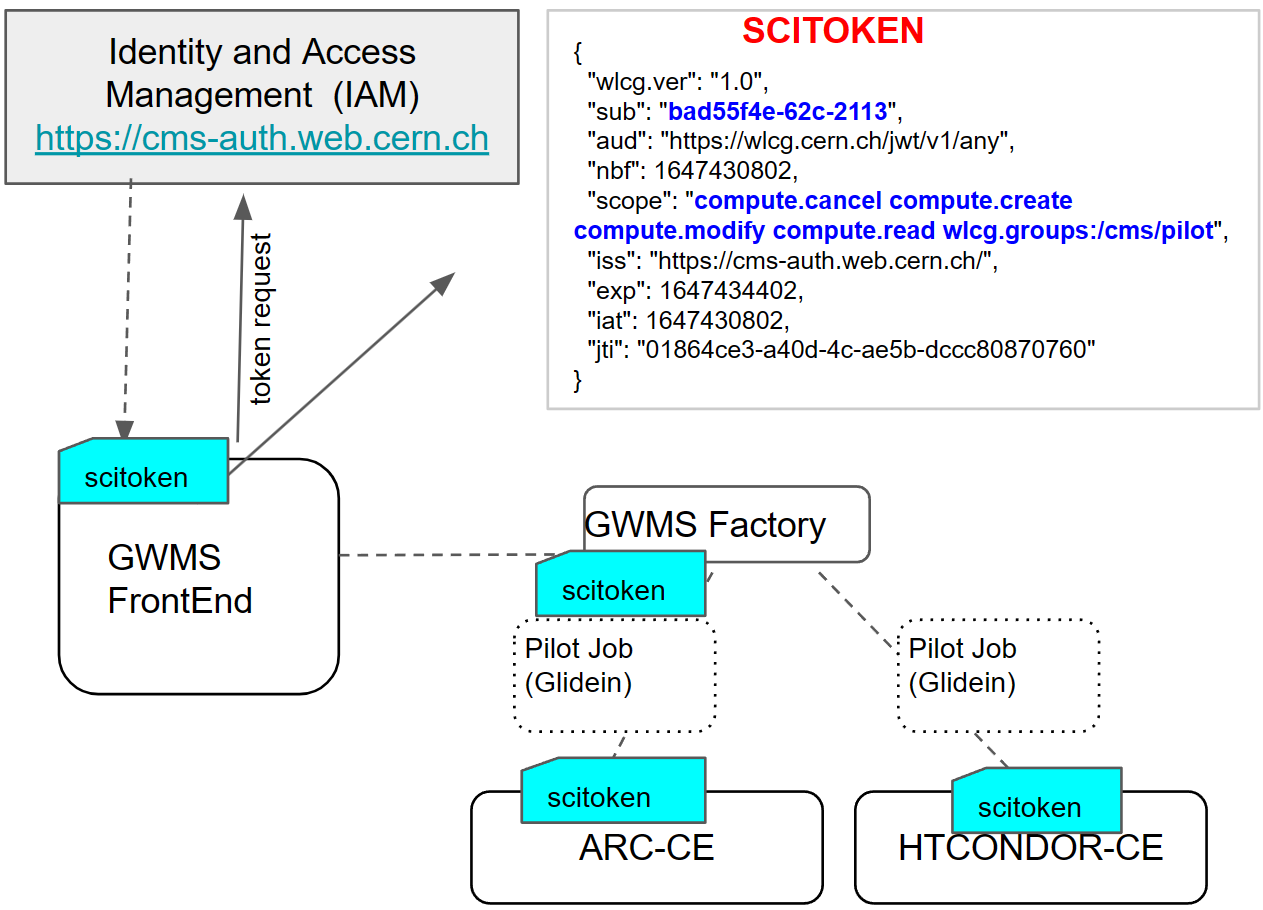}
\end{center}
\caption{The use of Scitokens in GlideinWMS to access Grid sites. Tokens are fetched from the CMS IAM instance by the Frontend and then propagated to the Factory to authorize pilot submission to CEs.}
\label{fig:SI_externals}
\end{figure}

Across the sites, two distinct Compute Element (CE) technologies are in use. HTCondor token submission has been successfully implemented across all sites, encompassing both OSG and EGI sites, with the final site completing this transition in January 2023. Scitokens are used to contact Grid sites as explained in Figure \ref{fig:SI_externals}. In contrast, the use of ARC-CEs for submission involved an LDAP interface and GSI proxies, but this approach has been deprecated in the HTCondor 10 series, which the GlideinWMS factory use to submit pilot jobs to CEs. To adapt to this change, a new REST interface has been made available, offering a more modern method for ARC-CEs submission.

The current focus within factory operations centers on enhancing ARC-CE submission processes. A campaign is underway to collaborate with sites, encouraging the adoption of the ARC-CE REST interface while configuring them to accept tokens. Rigorous testing in the Integration Test Bed (ITB) environment has been conducted to minimize any potential disruptions to working sites. However, it's worth noting that 12 out of 45 CEs (or 5 out of 25 sites) encountered issues with Scitoken access during this transition, which is an ongoing area of attention.

To ensure a seamless transition, a split configuration strategy has been implemented. The factory deployed at the University of California San diego (UCSD) utilized the new REST interface with tokens, while the CERN factory used the older LDAP interface. This approach guaranteed a fallback to proxy for pilot submission, preserving operational stability. Access to ARC-CE sites is currently performed using the REST interface with x509 certificates for both factories with plans to start using Scitokens for authentication in the UCSD factory.

While the transition has been largely successful, a few issues have been identified during scale usage. Notably, there have been instances of a memory leak in the arc\_gahp process on RHEL7, certain jobs remaining in a "run" state indefinitely under specific circumstances, and submission timeouts due to condor request serialization have been observed. All these issue are being addressed by the HTCondor development team.

\section{Conclusions and Future Works}

In Spring 2022, a significant milestone was achieved as all internal components of the CMS Submission Infrastructure successfully transitioned to IDTokens. This marked a crucial step towards modernizing and enhancing the security and authentication mechanisms within the CMS ecosystem. Additionally, by January 2023, communication to all HTCondor-CEs used by CMS has been migrated to Scitokens, while the submission to ARC-CE is now performed using the new REST interface plus X509 proxies deprecating the old ARC-CE LDAP interface.

The introduction of IDTokens brought about improved secret management practices. While GSI secrets required minimal maintenance due to their one-certificate-per-host approach, IDTokens demand a more rigorous approach to ensure their secure generation, storage, and distribution to their intended users. The adoption of Teigi for secret storage was a notable development in this regard, contributing to enhanced security and efficiency.

Looking ahead, several important tasks and initiatives remain on the horizon. The primary focus will be on completing the token transition by cleaning the HTCondor configurations from the GSI fallback in the global pool. Additionally, plans are in motion to start adopting tokens to access ARC-CEs. Furthermore, the implementation of security "drill exercises" will be essential to continually assess and strengthen the security posture of the infrastructure, ensuring its resilience in the face of emerging threats and challenges. 

\section*{Acknowledgements}
This work was partially supported by the Spanish Ministry of Science and Innovation under grants PID2019-110942RB-C21, PID2019-110942RB-C22 and PID2020-113807RA-I00, which include FEDER funds from the European Union, and by the US National Science Foundation under Grant No. 2121686


\begin{thebibliography}{}
%
%

\bibitem{cms} CMS Collaboration, The CMS experiment at the CERN LHC, 
JINST 3 (2008) S08004, doi:10.1088/1748-0221/3/08/S08004.
\bibitem{htcondor} The HTCondor Software Suite public web site, \url{https://research.cs.wisc.edu/htcondor/index.html}.
\bibitem{wlcg} The Worldwide LHC Computing Grid \url{http://wlcg.web.cern.ch}.
\bibitem{wlcgtimeline} WLCG Token Transition Timeline, \url{https://zenodo.org/record/7014668#.ZBLfC4DMKw5}.
\bibitem{htcondorce} B. Bockelman et al. ``Commissioning the HTCondor-CE for the Open Science Grid'',  J. Phys.: Conf. Ser. \textbf(664) 062003 (2015).
\bibitem{sicomplexity} A. Perez-Calero Yzquierdo et al. ``Evolution of the CMS Global Submission Infrastructure for the HL-LHC Era'', EPJ Web Conf. 245 (2020) 03016.
\bibitem{globalpool} J. Balcas et al. ``Using the glideinWMS System as a Common Resource Provisioning Layer in CMS'', J. Phys.: Conf. Ser. \textbf{664} 062031 (2015).
\bibitem{gwms} The Glidein-based Workflow Management System, \url{https://glideinwms.fnal.gov/doc.prd/index.html}.
\bibitem{pilots} I. Sfiligoi et al. ``The Pilot Way to Grid Resources Using glideinWMS'', 2009 WRI World Congress on Computer Science and Information Engineering, Los Angeles, CA, USA, 2009, pp. 428-432, doi: 10.1109/CSIE.2009.950.
\end{thebibliography}
\end{document}